\documentclass[twocolumn,showpacs,prl]{revtex4}
\usepackage{graphicx}
\usepackage{amssymb}

\def\fig#1{Fig.\,\ref{#1}}

\begin{document}

\title{Energy absorption of xenon clusters in helium nanodroplets
  under strong laser pulses}

\author{Alexey Mikaberidze}
\author{Ulf Saalmann}
\author{Jan M. Rost}
\affiliation{Max Planck Institute for the Physics of Complex Systems\\
N\"{o}thnitzer Stra{\ss}e 38, 01187 Dresden, Germany}

\begin{abstract}\noindent
Energy absorption of xenon clusters embedded in helium
nanodroplets from strong femtosecond laser pulses is studied
theoretically.  
Compared to pure clusters we find earlier and more efficient
energy absorption in agreement with experiments. 
This effect is due to resonant absorption of the helium
nanoplasma whose formation is catalyzed by the xenon core.
For very short double pulses with variable delay both plasma
resonances, due to the helium shell and the xenon core, are
identified and the experimental conditions are given which
should allow for a simultaneous observation of both of them. 
\end{abstract}

\maketitle

\noindent
Atomic clusters couple very efficiently to laser light pulses 
\cite{sasi+06,krsm02,po01}. 
Through irradiation with a strong laser pulse, typically a
nanoplasma is formed which can absorb resonantly energy if
its eigenfrequency matches the laser frequency \cite{dido+96}. While
it is generally agreed by now that collective electron oscillation
resonant with the laser frequency is the most efficient way to
transfer energy from the laser pulse into the cluster
\cite{dido+96,kosc+99,lajo99,saro03,sa06}, subtle effects continue to be
discussed \cite{juge+04,kuba06,fedo+07}.
On the other hand intriguing and so far not well understood
phenomena have been observed in composite clusters illuminated by
strong laser pulses, such as enhancement of X-ray production in
water-doped clusters \cite{jhma+05,jhsh+06} and the earlier
resonant energy absorption for clusters embedded in helium droplets
\cite{dote+03}.
Growing clusters in helium droplets has proven to be an elegant
alternative to their production in supersonic beams
\cite{lesc+95,dodi+01,tist07}. In contrast to spectroscopic
applications \cite{stle06}, where the role of the helium droplet is merely
to isolate and cool the embedded species, helium embedding may
significantly influence cluster dynamics when a strong laser pulse is
applied \cite{dote+03}.

Here, we investigate this influence theoretically by exposing a xenon
cluster (100 atoms) embedded in a helium droplet of up to 5000 atoms to a
single laser pulse. It is taken sufficiently long (100\,fs) for an
expansion of the cluster to roughly double its original size. We find
the usual resonant absorption when the eigenfrequency of the expanding
xenon cluster matches the laser frequency. However, in agreement with
the experiment \cite{dote+03}, earlier in the pulse we see another
resonance which dominates the energy absorption. Therefore, higher
cluster charges and energy absorption are observed in a helium-embedded
cluster compared to the pure one. This suggests that the resonance
earlier in the laser pulse found for helium embedded lead clusters
\cite{dote+03} is not the same one as seen for the pure lead cluster,
simply shifted to earlier times due to the helium environment. Rather, it
is the resonance of the embedding helium droplet and for some reason to be
clarified the resonance of the xenon core is not visible.

To understand better the energy absorption, we have used the
double pulse scheme and replaced the 100\,fs pulse by two
time-delayed very short pulses of 10\,fs duration, during which
the cluster expansion is negligible. As we
will show the cluster parameters as well as those of the laser
have to be chosen very carefully to see both resonances in one
experiment.

We use a classical molecular dynamics approach with tree-code
techniques \cite{bahu86,sasi+06} to follow the cluster dynamics. The
initial positions of the xenon atoms are chosen according to the lowest
energy configuration \cite{wado+07}, while helium atoms are placed
randomly so that the droplet has the density of bulk liquid helium 
($\rho=0.02185$\,\AA$^{-3}$). As ionization mechanism we assume barrier
suppression which dominates at the high laser intensities considered
here \cite{lajo99}. This means that due to electric fields created
by surrounding charges and the laser the potential barrier is bent
down sufficiently to release classically a bound electron from its
mother ion into the cluster environment \cite{sa06}. Such electrons
are called ``inner-ionized'' and their number equals the positive
background charge $Q(t)$ created. Some of them gain enough energy to
leave the cluster, they become ``outer-ionized''. Those, which are
inner-ionized but remain in the cluster, we call ``quasi-free''.
They form the nanoplasma \cite{dido+96} which plays a crucial
role for the energy absorption in the cluster \cite{sasi+06}. 
All the laser pulses used have 780\,nm wavelength. For clarity we
will restrict ourselves to a xenon cluster of 100 atoms embedded in
helium droplets of 300, 500, 1000 and 5000 atoms, respectively.

\begin{figure}
 \includegraphics[width=\columnwidth]{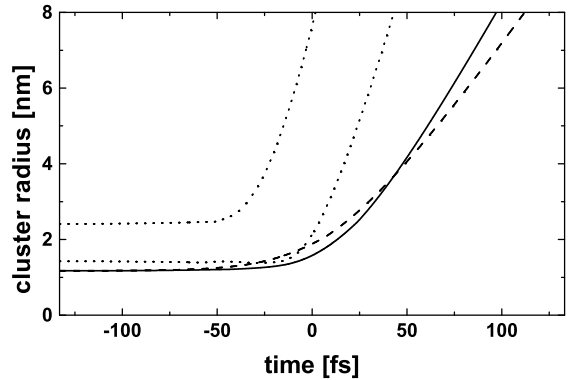}
 \caption{\label{fig:XeHe-radius}
 Cluster radii as functions of time for the 
 helium droplet in Xe$_{100}$@He$_{1000}$ (dotted, inner and 
 outer edge), for the xenon cluster in Xe$_{100}$@He$_{1000}$ 
 (solid), and for the bare Xe$_{100}$ cluster (dashed).
 }
\end{figure}
\begin{figure}
 \includegraphics[width=\columnwidth]{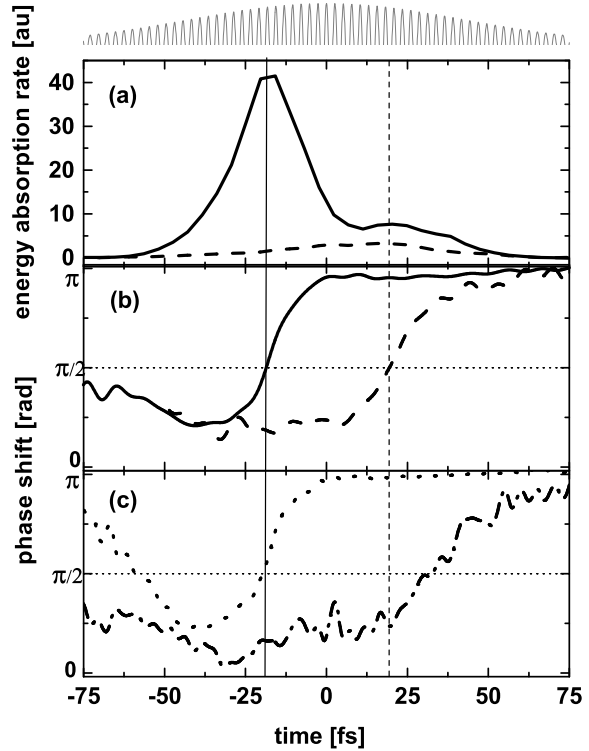}
 \caption{\label{Fi:pshifts-hilo} Energy absorption rate (a), and phase
 shift of CM oscillations of the quasi-free electrons with respect to
 laser field (b) for Xe$_{100}$@He$_{1000}$ (solid) and Xe$_{100}$
 (dashed), respectively.
 Phase shifts for the the embedded system Xe$_{100}$@He$_{1000}$
 (c) from electrons in the helium shell  (dotted) and in the
 xenon core (dash-dotted) only.  
 The laser pulse with a peak intensity of 
 $I=3.51 \times 10^{14}$\,W/cm$ ^2$,  a duration of 100\,fs
 and a wavelength of $\lambda = 780 $\,nm  is shown on
 top of the figure.  
 }
\end{figure}

%
First, we will discuss the energy absorption of the embedded cluster
from a single laser pulse of 100\,fs duration (FWHM) with the
intensity $I=3.51 \times 10^{14}$\,W/cm$^2$. Under this laser pulse,
the cluster is ionized and expands. Every helium atom is doubly ionized,
while the average charge of xenon is around 11+ per ion by the end of the
pulse. The expansion is much faster for the light helium ions than for
the heavier xenon ions (see \fig{fig:XeHe-radius}). While the cluster
expands, the rate of energy absorption changes dramatically, reaching
two maxima, as can be seen from \fig{Fi:pshifts-hilo}a. The first
maximum is much higher than the second one. Comparison with the pure
cluster (dashed line) indicates that the second maximum is caused by
the resonant absorption due to the expanding xenon cluster: At resonance
the xenon cluster has reached a critical radius
$R(t^\star)=R_\mathrm{crit}$ so that the eigenfrequency of the
quasi-free electrons  $\Omega(t)= \sqrt{Q(t)/R^3(t)}$ matches
the laser frequency $\Omega(t^\star)=\omega$ \cite{saro03}. 
A clearer indication for resonant absorption than the
eigenfrequency is the time dependent phase shift $\phi(t)$ of the
periodic center of mass (CM) motion of the quasi-free electrons with
respect to the driving laser field. At resonance the phase shift
is  $\phi(t^\star)=\pi/2$ \cite{saro03}.
The corresponding times $t^\star$ are indicated by the two
vertical lines  in \fig{Fi:pshifts-hilo}b for the embedded
(solid) and the pure (dashed) cluster.  

Indeed, also for the first and the dominant resonance feature, the
phase shift of the CM motion of \emph{all} quasi-free electrons
passes $\pi/2$ at the time of maximum absorption rate. Excluding
the quasi-free electrons inside the xenon core of the cluster gives
the same result (dotted curve in \fig{Fi:pshifts-hilo}c) which
clarifies that it is the period of the electronic CM motion in the
extended potential of the helium ions which coincides with the
laser period. This result also explains why the early resonance
is much stronger than the second one. It is simply because there
are many more helium ions than xenon ones so that many more electrons
participate in the resonant absorption in helium shell than in xenon
core. Note, that for a larger helium droplet of 5000 atoms the second
resonance is hidden by the first one and not seen in the energy
absorption rate.

These findings are consistent with the earlier absorption in the
embedded cluster compared to the pure one found in the experiment
\cite{dote+03}. However, in contrast to the two resonances we have
identified, only one early in the laser pulse was seen in the
experiment using a pump-probe double pulse scheme. Therefore, we will
analyze the effect of the double pulse with variable delay on the same
system as before in the following.

\begin{figure}
  \includegraphics[width=\linewidth]{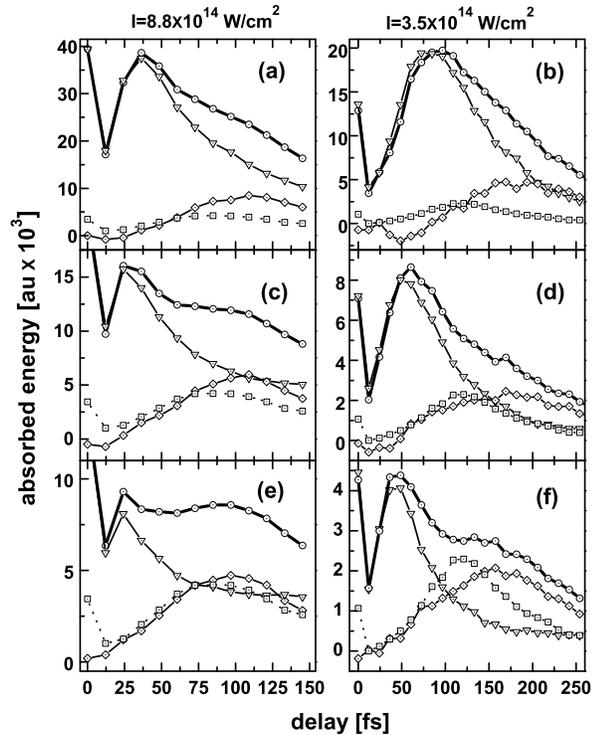}
  \caption{\label{Fi:absen-vs-delay} Absorbed energy as a function 
  of delay between pump and probe pulse
  for Xe$_{100}$@He$_{N}$ ($\circ $), for xenon atoms 
  fixed ($\triangledown $), for bare Xe$_{100}$ clusters ($\Box $, 
  dotted line) and finally, the difference in energy absorption for 
  helium-embedded xenon clusters
  with moving and fixed xenon atoms ($\Diamond $). The panels (a) 
  and (b) are for $N=1000$ helium atoms, (c) and (d) for $N=500$ 
  and (e) and (f) for $N=300$, respectively.
  }
\end{figure}

We have used two identical pulses with a duration of 10\,fs (FWHM)
and a delay from 10 to 250\,fs. Results for two different
intensities of the pulses ($I=8.8 \times 10^{14}$\,W/cm$^2$ and
$I=3.5 \times 10^{14}$\,W/cm$^2$) are shown in
\fig{Fi:absen-vs-delay}. Surprisingly, for the same cluster as
before (Xe$_{100}$@He$_{1000}$), one only sees the early resonance
(\fig{Fi:absen-vs-delay}a) in qualitative agreement with the
experiment \cite{dote+03} for which we conclude that there the helium
resonance was observed.

For the higher intensity (\fig{Fi:absen-vs-delay}a) the resonance
peak appears to be rather asymmetric having a shoulder towards
longer delays where the second resonance due to xenon should be but
is apparently masked by the helium resonance. There are several ways
to check this assumption: The one which can also be realized
experimentally is to increase the number of xenon atoms relative to
the helium atoms. We did so by reducing the number of helium atoms.
Indeed, for 500 helium atoms the second resonance appears already for
the higher intensity, and for 300 helium atoms, the later resonance is
visible at both laser intensities. This shows, that the parameters
of the laser have to be carefully adopted for the embedded cluster
under investigation in order to get the full information on energy
absorption from pump-probe schemes.

To further elucidate the nature of the second smaller energy
absorption peak, we have fixed the xenon atoms in space. This is, of
course, only possible in a calculation. For such a situation we can
exclude resonant absorption by the xenon core since the ion charge
density there will be too high for matching the corresponding
eigenfrequency with the laser frequency (triangles in
\fig{Fi:absen-vs-delay}). Consequently, under none of the parameter
combinations energy absorption for fixed xenon shows a second peak in
\fig{Fi:absen-vs-delay}. To demonstrate that the xenon resonance
nevertheless exists, we have constructed an artificial absorption
curve (squares) by subtracting the result for fixed xenon from the full
dynamical absorption (circles). Now, again for all parameter
combinations, the first resonance due to the helium ions has disappeared.
This indicates that it has not been affected by the presence of the
fixed xenon atoms and ions. On the other hand, the second resonance due
to xenon clearly shows up in the difference curve, revealing that it is
present but buried under the helium signal in the full dynamical energy
absorption (circles) but of course by construction absent in the fixed
xenon absorption (triangles). Moreover, this difference curve bares
similarity with the energy absorption for a pure xenon cluster, where the
general trend is that the xenon resonance is stronger and appears later
in the embedded cluster. It is stronger due to the larger number of
quasi-free electrons participating in the resonant absorption. They
also give rise to a slightly higher ion charge $Q(t)$ of the embedded
xenon cluster. Consequently, a longer expansion time is needed to reach the
critical ion charge density corresponding to resonance.

\begin{figure}
 \includegraphics[width=\columnwidth]{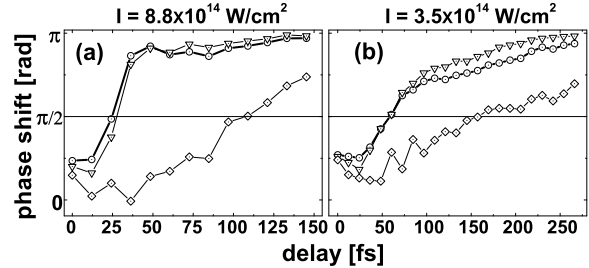}
 \caption{\label{Fi:pshifts}Phase shift of the CM oscillation of
 quasi-free electrons with respect to the driving field taken at the
 maximum of the second pulse as a function of delay between pulses.
The phase shift of all quasi-free electrons
 ($\circ $), of quasi-free electrons in the helium shell only
 ($\triangledown $) and in the xenon core
 ($\Diamond $) only is shown for Xe$_{100}$@He$_{500}$.}
\end{figure}%
We finally take a look at the phase shift of the quasi-free
electron CM motion versus the laser driving in \fig{Fi:pshifts}
for the case of 500 helium atoms. We have separated the quasi-free
electrons contributing to the helium resonance and to the xenon one by
spatial selection of the respective electrons in the cluster. The
result confirms our picture: The second weaker absorption peak is
due to the electrons near the xenon core. Their collective phase
shift $\phi(t)$ passes $\pi/2$ at the right time of about 100\,fs
and 160\,fs for the two laser intensities, respectively.

From these results one could get the impression, that in a composite
cluster of two atomic species two resonances occur related to the
respective two resonance frequencies, or specifically in our case: we
observe the helium resonance of the droplet almost not influenced by the
xenon core. This is, however, by no means true: We have verified that
without the xenon core, helium is not ionized at all for the laser
intensities used. This is clear recalling that the intensities are
not large enough to ionize helium by barrier suppression, while
multiphoton ionization requires some 20 photons and is very unlikely.
Moreover, when the laser intensity is increased so that barrier
suppression ionization of helium becomes possible, our calculations show
that resonant absorption is of minor importance (less than 20\%
of the total absorbed energy) in contrast to a bare cluster. The
reason is that double ionization follows single ionization immediately,
as soon the laser intensity is enough for single ionization of helium,
due to strong fields
that build up in the cluster environment. After that, there is
only a small number of quasi-free electrons left in the cluster, which
become outer ionized as the cluster approaches the plasma resonance. On
the other hand, a xenon core (or similar) is more easily ionized and
first drives the electrons, removed from helium, to the center of the
cluster instead of loosing them. Thus, it is only the composite
cluster (xenon cluster in a helium droplet) that exhibits the earlier resonance
leading to very strong energy absorption. Neither the helium nor the 
xenon cluster by itself has this property.

To summarize, we have shown that a helium embedding strongly influences
the dynamics of rare gas clusters illuminated by strong laser pulses.
For both, single and double laser pulses, the helium droplet leads to the
appearance of an additional plasma resonance, occurring earlier in the
laser pulse. Its weight relative to the plasma resonance in the xenon
cluster core depends on the size of the droplet compared to the size
of xenon cluster and on the laser intensity. This finding should allow
to choose the parameters such that an observation of both resonances
in one double-pulse experiment can be realized. The dominant
resonance in helium droplet occurs earlier because light helium ions explode
faster than xenon ions. Due to this additional resonance and also due to
the larger number of quasi-free electrons the helium embedding increases
the energy absorption of the cluster. This indicates that helium
embedding is quite dangerous if one is interested in the 
properties of the embedded clusters \cite{dofe+05}, since the
dynamics of the helium 
droplet dominates the absorption properties of the composite
cluster in most cases. 

Furthermore, based on the results presented we conjecture that
significant resonant absorption in a helium droplet is only possible in a
composite cluster unless the helium droplet is very large. From the
perspective of xenon, however, the helium embedding makes the absorption
slightly stronger since more electrons, also from helium, are available.
Finally, our results suggest that it may be well possible to obtain,
e.g., very fast electrons or X-rays dominantly of a certain wavelength
by specifically choosing a certain composite cluster in combination
with a suitable dual laser pulse. Studies in this direction are under
way.

We thank Ionu\c{t} Georgescu and Christian Gnodtke for helpful
discussions and the International Max Planck Research School
``Dynamical Processes in Atoms, Molecules and Solids'' for
financial support.

\end{document}